\documentclass[aip,reprint,amsmath,amssymb]{revtex4-1}

\usepackage{graphicx}
\usepackage{dcolumn}
\usepackage{float}
\usepackage{bm}
\usepackage{amssymb}
\usepackage{lipsum}

\begin{document}

\title{Free energy and segregation dynamics of two channel-confined polymers of different length}

\author{James M. Polson} 
\affiliation{ Department of Physics, University of Prince Edward Island, 
550 University Avenue, Charlottetown, Prince Edward Island, C1A 4P3, Canada }
\author{Qinxin Zhu}
\altaffiliation{Current address: Department of Physics and Atmospheric Science,
 Sir James Dunn Building, 6310 Coburg Road, Halifax, Nova Scotia, Canada, B3H 4R2}
\affiliation{ Department of Physics, University of Prince Edward Island,
550 University Avenue, Charlottetown, Prince Edward Island, C1A 4P3, Canada }

\date{\today}

\begin{abstract}
Polymers confined to a narrow channel are subject to strong entropic forces that tend to drive 
the molecules apart.  In this study, we use Monte Carlo computer simulations to study the 
segregation behavior of two flexible hard-sphere polymers under confinement in a cylindrical channel. 
We focus on the effects of using polymers of different lengths. We measure the variation of 
the conformational free energy, $F$, with the center-of-mass separation distance, $\lambda$.
The simulations reveal four different separation regimes, characterized by different 
scaling properties of the free energy with respect to the polymer lengths and the 
channel diameter, $D$. We propose a regime map in which the state of the system is
determined by the values of the quantities $N_2/N_1$ and $\lambda/(N_1+N_2)D^{-\beta}$,
where $N_1$ and $N_2$ are the polymer lengths, and where $\beta\approx 0.64$.
The observed scaling behavior of $F(\lambda)$ in each regime is in reasonable agreement 
with predictions using a simple theoretical
model.  In addition, we use MC dynamics simulations to study the segregation dynamics 
of initially overlapping polymers by measurement of the incremental mean first-passage 
time with respect to $\lambda$. For systems characterized by a wide range of $\lambda$ in 
which a short polymer is nested within a longer one, the segregation dynamics are close to that 
expected for two noninteracting 1D random walkers undergoing unbiased diffusion. When the 
free energy gradient is large segregation is rapid and characterized by out-of-equilibrium 
effects. 
\end{abstract}

\maketitle

\section{Introduction}
\label{sec:intro}

The overlap of two polymer chains leads to a reduction in the total conformational entropy
and thus an increase in the free energy. Often described as an ``entropic force'', the gradient 
in the free energy tends to drive the polymers apart to reduce the loss in entropy.%
\cite{grosberg1982polymeric,dautenhahn1994monte} When the polymers are confined to 
anisometric spaces with dimensions small compared to the bulk radius of gyration this 
inter-polymer entropic repulsion is generally enhanced.\cite{daoud1977statistics,jun2006entropy} 
A notable example is polymer confinement in nano-channels, where the entropic force causes the 
polymers to segregate along the axis of the channel. This process has been the subject of numerous 
theoretical and computer simulation studies in recent years.\cite{jun2006entropy,teraoka2004computer,%
jun2007confined,arnold2007time,jacobsen2010demixing,jung2010overlapping,jung2012ring,jung2012intrachain,%
liu2012segregation,dorier2013modelling,racko2013segregation,shin2014mixing,minina2014induction,%
minina2015entropic,chen2015polymer,polson2014polymer,du2018polymer,polson2018segregation,%
nowicki2019segregation,%
nowicki2019electrostatic} Most of these studies have examined the segregation behavior of
flexible linear polymers,\cite{jun2006entropy,teraoka2004computer,jun2007confined,%
arnold2007time,jacobsen2010demixing,jung2010overlapping,jung2012intrachain,%
liu2012segregation,racko2013segregation,minina2014induction,polson2014polymer,du2018polymer} 
though some have also investigated the behavior of ring polymers.\cite{jung2012ring,%
dorier2013modelling,shin2014mixing,minina2015entropic,chen2015polymer,polson2018segregation}
In addition, the effects of of bending rigidity,\cite{racko2013segregation,polson2014polymer} 
macromolecular crowding,\cite{shin2014mixing,chen2015polymer,polson2018segregation}
and electrostatics\cite{nowicki2019segregation,nowicki2019electrostatic} on the
segregation dynamics and thermodynamics have been examined in detail.

A key motivation for theoretical studies of entropy-driven separation of confined polymers
is to elucidate the role that entropy plays in the process of chromosome segregation
in self-replicating bacteria.\cite{jun2006entropy,jun2010entropy,diventura2013chromosome,%
youngren2014multifork,mannik2016role} Experiments have shown that the newly formed arms of
a chromosome segregate as duplication proceeds while the terminus region remains at the
center of the nucleoid.\cite{badrinarayanan2015bacterial} Modelling the chromosome as a polymer
chain, Jun and Mulder showed that entropic demixing alone could account for such behavior.%
\cite{jun2006entropy} Though some experimental studies have yielded results that are hard
to reconcile with a purely entropic segregation mechanism,\cite{fritsche2011model,%
benza2012physical,yazdi2012variation,fisher2013four,diventura2013chromosome,kuwada2013mapping}
more recent studies have reported results consistent with this hypothesis.\cite{diventura2013chromosome,%
mannik2016role,cass2016escherichia,wu2019geometric,elnajjar2020chromosome,japaridze2020direct}
Though a full understanding of the process is lacking\cite{badrinarayanan2015bacterial} it 
appears likely that entropic forces contribute to driving segregation, perhaps in conjunction with 
other mechanisms.\cite{diventura2013chromosome,mannik2016role,hofmann2019self}

Another motivation for theoretical studies of separation of confined polymers arises from
recent advances in nanofluidic technology that facilitate the study {\it in vitro} of the 
physical behavior of multiple polymers confined to small spaces. A recent study by Capaldi 
{\it et al.} used fluorescence microscopy to study the dynamics and organization of two
differentially stained $\lambda$-DNA molecules confined to rectangular nano-cavities.%
\cite{capaldi2018probing} They also examined the confinement of a single $\lambda$-DNA 
molecule with a plasmid, which is a small circular DNA molecule typically found in bacteria
outside the nucleoid.  Such systems provide simple models of prokaryotic organisms that may 
help clarify the role of entropy as a demixing mechanism. They may also be useful for
understanding the organization of multiple chromosomes inside a eukaryotic cell nucleus. 
In addition to contributing to understanding biological processes,  insight provided by 
such experiments has important technological relevance. For example, the mixing/partitioning 
properties of multiple chains confined to channels or cavities may affect the performance of 
nanofluidic devices used for polymer manipulation and separation.\cite{han2000separation,%
turner2002confinement} 

Our recent work on confinement effects of multiple-polymer systems has focused on the calculation 
of the overlap free energy of polymers inside channels and elongated cavities using Monte Carlo 
simulation methods.\cite{polson2014polymer,polson2018segregation} Specifically, we have examined 
the variation of the free energy with respect to the polymer separation distance as well as other 
variables.  Such free energy functions are closely connected to the polymer mixing/segregation 
tendencies and the separation dynamics. As a measure of separation, we use the distance between the 
polymer centers of mass along the channel, in part because such a quantity is experimentally
measurable.\cite{capaldi2018probing} Our goal has been to fully characterize these functions 
with respect to the key system parameters. These parameters include the confinement dimensions, 
polymer length, bending rigidity, the density of crowding agents, and the chain topology. We 
have shown that the scaling properties of the free energy functions are reasonably consistent 
with predictions based on standard scaling theories in polymer physics. Thus far, we have considered 
only systems characterized by a high degree of symmetry, e.g. two identical linear or ring 
polymers confined to a cylindrical channel.  

The purpose of the present study is to extend this work to examine the behavior of systems 
characterized by an asymmetry in the lengths of the polymers. This is expected to be relevant
to the operation of nanofluidic devices for manipulation and analysis of DNA and other polymers. 
As in our other studies on this topic,\cite{polson2014polymer,polson2018segregation} we examine
the behavior of two self-avoiding hard-sphere chains confined to a hard-walled channel.
In addition to measuring and characterizing the free energy functions, we also use MC dynamics 
simulations to investigate the segregation dynamics and interpret the results in light of the 
calculated free energies.  The free energy results are analyzed and interpreted using standard 
scaling theory methods.  Generally, the results are in semi-quantitative agreement with most 
predictions, though scaling exponents deviate slightly from theoretical values due to 
finite-size effects and deficiencies in the theoretical model.  In addition, the segregation 
dynamics behavior align with expectations based on the equilibrium free energy functions, 
though out-of-equilibrium effects are evident at separation distances corresponding to 
strong entropic forces. 

\section{Model}
\label{sec:model}

We examine a system of two flexible polymer chains confined to an infinitely long channel. Each 
polymer is modeled as freely-jointed chain of hard spheres, where each spherical monomer has a
diameter $\sigma$.  The pair potential for non-bonded monomers is thus $u_{\rm{nb}}(r)=\infty$
for $r\leq\sigma$, and $u_{\rm{nb}}(r)=0$ for $r>\sigma$, where $r$ is the distance between
the centers of the monomers. Pairs of bonded monomers interact with a potential
$u_{\rm{b}}(r)= 0$ if $0.9\sigma<r<1.1\sigma$ and $u_{\rm{b}}(r)= \infty$, otherwise.
The polymer length is given in terms of the number of monomers of each polymer, $N_1$ and
$N_2$, where we choose $N_2\leq N_1$. 

The polymers are confined to a hard-wall cylindrical channel that is aligned along the $z$-axis. 
The channel has an effective diameter $D$ such that each monomer interacts with the wall of 
the tube with a potential $u_{\rm w}(r) = 0$ for $r<D$ and $u_{\rm w}(r) = \infty$ for $r>D$, 
where $r$ is the distance of the monomer center from the central axis of the cylinder. Thus, 
$D$ is the diameter of the cylindrical volume accessible to the centers of the monomers, and
the actual diameter of the cylinder is $D+\sigma$. 

\section{Methods}
\label{sec:methods}

In most simulations we measure the variation of the free energy with $\lambda$, the 
distance between the polymer centers of mass along the cylinder axis.  We employ Monte Carlo 
simulations with the Metropolis algorithm together with the self-consistent histogram (SCH) 
method.\cite{frenkel2002understanding} To implement the SCH method, we carry out many 
independent simulations, each of which employs a unique ``window potential'' of the form:
\begin{eqnarray}
{W_i(\lambda)}=\begin{cases} \infty, \hspace{8mm} \lambda<\lambda_i^{\rm min} \cr 0,
\hspace{1cm} \lambda_i^{\rm min}<\lambda<\lambda_i^{\rm max} \cr \infty,
\hspace{8mm} \lambda>\lambda_i^{\rm max} \cr
\end{cases}
\label{eq:winPot}
\end{eqnarray}
where $\lambda_i^{\rm min}$ and $\lambda_i^{\rm max}$ are the limits that define the range
of $\lambda$ for the $i$-th window.  Within each window of $\lambda$, a probability
distribution $p_i(\lambda)$ is calculated in the simulation. The window potential width,
$\Delta \lambda \equiv \lambda_i^{\rm max} - \lambda_i^{\rm min}$, is chosen to be
sufficiently small that the variation in $F$ does not exceed a few $k_{\rm B}T$.
The windows are chosen to overlap with half of the adjacent window, such that
$\lambda^{\rm max}_{i} = \lambda^{\rm min}_{i+2}$.  The window width was typically in the range
$\Delta \lambda = 2\sigma-4\sigma$. The SCH algorithm was employed to reconstruct the unbiased
distribution, ${\cal P}(\lambda)$, from the $p_i(\lambda)$ histograms. The free energy
follows from the relation $F(\lambda) = -k_{\rm B}T\ln {\cal P}(\lambda)+{\rm const}$.
We choose the constant such that $F(\lambda=\infty)=0$.

In some simulations we consider polymers with centers of mass that overlap along the channel
and measure the variation of the free energy with $\zeta\equiv X_1-X_2$, where $\zeta$ is
difference in the spans of the two polymers along the channel, $X_1$ and $X_2$. The SCH method
is implemented for $\zeta$ in exactly the same manner as for $\lambda$. Specifically, we use
window potentials of the form of Eq.~(\ref{eq:winPot}) for $\zeta$ such that trial MC moves
that yield values of $\zeta$ outside the range of the window are rejected. The resulting
collection of probability distributions $p_i(\zeta)$ are used with the SCH algorithm to 
reconstruct the underlying distribution ${\cal P}(\zeta)$ that yields the free energy,
$F(\zeta)=-k_{\rm B}T\ln{\cal P}(\zeta)$.

Polymer configurations were generated by carrying out single-monomer moves using a combination of
translational displacements and crankshaft rotations. In addition, whole-polymer displacements
and reptation moves of each polymer along the channel axis were also employed to increase the 
efficiency of sampling $p_i(\lambda)$.  Each trial move was rejected if it resulted in overlap
between particles or between a particle and a confinement surface, or if it led to a violation
of the bonding constraints; otherwise it was accepted. Measurements of the various quantities were
carried out every 100 MC cycles over a run time whose duration was ${\cal O}(10^7-10^8)$ MC
cycles. A single MC cycle corresponds to attempting each of the following coordinate changes once,
on average: (1) movement of a single randomly chosen monomer by translation or crankshaft rotation;
(2) whole-polymer displacement of a randomly selected polymer along the channel by changing all 
the $z$-coordinates by the same random value; (3) reptation at a randomly chosen end of a randomly 
selected polymer. After initializing the particle positions, the simulation proceeded through an 
equilibration period prior to sampling data.  Since the polymer configurations were initially 
chosen for convenience to be linear and aligned along the channel, such an equilibration period 
is necessary for the system to relax out of this highly artificial initial state. As an 
illustration, for two identical polymers of length $N_1=N_2=300$ in a $D$=6 cylindrical channel, 
the system was equilibrated for $1\times 10^7$ MC cycles, following which a production run of 
$4\times 10^7$ MC cycles was carried out. Equilibration and production run times were typically 
chosen to be greater for longer polymer systems.  

In addition to free energy calculations, Monte Carlo dynamics simulations were used to study
the segregation dynamics of the two confined polymers. The MC dynamics method is chosen here
since the Langevin dynamics method is inapplicable to molecular models with discontinuous
potentials. Provided the random trial moves are local and do not lead to chain crossing, 
the random motions of individual monomers can be viewed as discretized realizations of a
stochastic dynamical process such as that described by a Langevin equation.\cite{landau2009guide}
Consequently, for this model diffusion obeys Rouse scaling of the diffusion coefficient, i.e.,
${\cal D}\propto 1/N$. Though MC dynamics does not provide an absolute time scale or enable 
probing the short-time dynamical behavior such as bond vibration, it can be effective in
determining scaling exponents associated with polymer dynamics on longer time scales.%
\cite{polson2013polymer} As in Ref.~\onlinecite{polson2014polymer} we choose to simulate polymer 
motion solely through random monomer displacement. The coordinates of a randomly chosen monomer 
were displaced by an amount $\Delta r_{\alpha}$ for $\alpha=x,y,z$. Each coordinate 
displacement was randomly chosen from a uniform distribution in the range $[-0.14\sigma,0.14\sigma]$. 
The acceptance criteria are the same as described above for the free energy calculations,
with the exception that the potential of Eq.~(\ref{eq:winPot}) is not applied.
The range chosen for the distribution of the single-monomer trial displacements yields
acceptance ratios slightly below 50\%, corresponding to efficient sampling of configurations 
during the simulation. In effect, this choice also calibrates the time scale. 
The polymer configurations were initially constrained to $\lambda = 0$ while the polymer 
chains were equilibrated. Following equilibration, the constraint was removed and the polymers 
underwent segregation along the channel. The main quantity we measure is the incremental 
mean first-passage time (IMFPT), $\tau$, defined as the first time the polymers are separated 
by a given value of $\lambda$. The IMFPT was originally introduced to characterize dynamics 
in polymer translocation simulations.\cite{deHaan2011incremental,deHaan2012using}
During the segregation process we measure the degree of overlap of the polymers as well 
as their extension lengths. We simulated typically 100--200 segregation events to calculate 
averages of these quantities.

\begin{figure}[!ht]
\begin{center}
\vspace*{0.2in}
\includegraphics[width=0.45\textwidth]{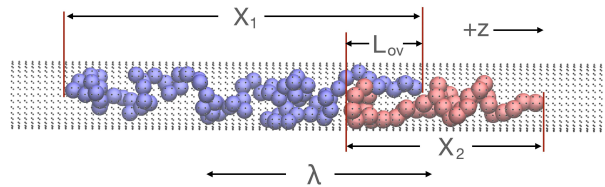}
\end{center}
\caption{Snapshot of two polymers with lengths $N_1$=80 and $N_2$=40 in a cylindrical
tube of diameter $D$=4. The polymer extension lengths, $X_1$ and $X_2$, the overlap 
distance, $L_{\rm ov}$, and the distance between polymer centers of mass, $\lambda$, 
are each labeled in the figure.}
\label{fig:illust}
\end{figure}

For both the free energy calculations and the MC dynamics simulations we calculate the
mean extension lengths, $X_1$ and $X_2$, of each polymer along the channel. We also measure
the mean overlap distance $L_{\rm ov}$, which is the range of positions along the channel 
occupied by part of the contours of both chains. These quantities are illustrated in 
Fig.~\ref{fig:illust}, which shows a snapshot of a system of two partially overlapping 
polymers. Note that $L_{\rm ov}$ is more generally defined as follows. With reference to the
illustration in Fig.~\ref{fig:illust}, it is the difference between the $z$ coordinate of 
the rightmost monomer of the longer (blue) polymer and the leftmost monomer of the shorter 
(red) polymer.  Clearly, when the polymers overlap, $L_{\rm ov}$ is the overlap distance 
shown in the figure.  However, if the polymers do not overlap, then $L_{\rm ov}<0$, and its 
magnitude is simply the distance along the $z$ axis between the nearest pair monomers, one 
on each polymer.

In the results presented below, distances are measured in units of $\sigma$, energy
is measured in units of $k_{\rm B}T$, and time is measured in MC cycles.

\section{Results}
\label{sec:results}

\subsection{General features of the free energy functions}
\label{subsec:overlap}

\begin{figure}[!ht]
\begin{center}
\vspace*{0.2in}
\includegraphics[width=0.45\textwidth]{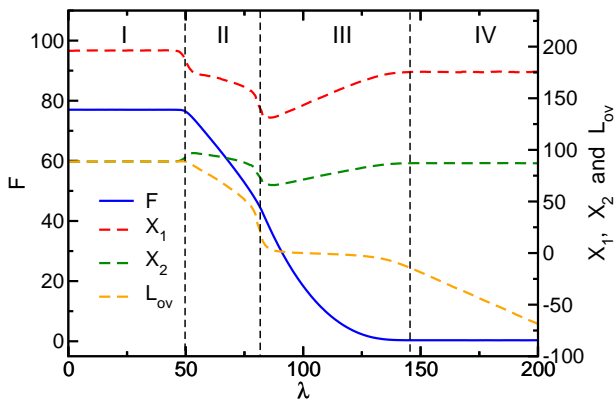}
\end{center}
\caption{Free energy vs $\lambda$ for polymers of length $N_1$=400 and $N_2$=200 
confined to a cylinder of diameter $D$=4. Overlaid on the graph are the extension
lengths of the polymers, $X_1$ and $X_2$, as well as the overlap distance, $L_{\rm ov}$,
as defined in the text. The vertical dashed lines denote approximate boundaries
of the four regimes described in the text.}
\label{fig:F_illust} 
\end{figure}

Let us first examine the general qualitative features of the free energy functions.
As an illustration, Fig.~\ref{fig:F_illust} shows the variation of the free energy 
with respect to the center-of-mass distance $\lambda$ for polymers of length $N_1$=400
and $N_2$=200 confined to a channel of diameter $D$=4. The graph also shows
the variation of the extension lengths, $X_1$ and $X_2$, as well as the overlap
distance, $L_{\rm ov}$, with $\lambda$. Four different regimes are labeled, each
corresponding to a physically distinct state. 

In regime~I at low center-of-mass
separation ($\lambda\lesssim$50), the free energy and the overlap and extension
lengths are all constant with respect to $\lambda$. This regime corresponds to the
``nesting'' of the shorter polymer within the longer polymer. In this case, changing
the center-of-mass distance changes neither extension length nor the overlap
length (which is equal to $X_2$). At $\lambda\approx 50$, the system undergoes
a transition to regime~II, which is characterized by a drop in $F$.  At the transition, 
the extension of the longer polymer abruptly decreases at the transition while that for 
the shorter polymer slightly increases.  This regime corresponds to partial overlap of 
the polymers along the channel (as illustrated in Fig.~\ref{fig:illust}). 
As $\lambda$ decreases, the overlap distance decreases, as does each of the extension 
lengths.

At $\lambda\approx 80$, the system undergoes transition into regime~III, where the
curvature of $F(\lambda)$ changes from positive to negative. Significantly, the polymers 
have essentially no longitudinal overlap ($L_{\rm ov}\approx 0$), and the extension lengths 
each increase with $\lambda$.  This regime corresponds to non-overlapping polymers that are 
longitudinally compressed and pressed against each other.  Varying $\lambda$ controls the 
degree of compression of the chains, each of which behaves like an entropic spring with
a free energy that rises rapidly with increasing compression.  

At $\lambda\approx 145$ the system 
undergoes a final transition to regime~IV, which corresponds to no physical contact between 
the polymers.  $L_{\rm ov}$ becomes increasingly negative with $\lambda$, indicating a growing 
space between the polymers.  As expected, the extension lengths and the free energy are invariant 
with respect to $\lambda$ here.  Qualitatively similar behavior for all four regimes was observed 
for polymers of equal length in Ref.~\onlinecite{polson2018segregation}, where it is described and 
explained in greater detail.  The key differences here are a wider nesting regime, unequal 
values of $X_1$ and $X_2$, and a slight increase in $X_2$ at the I-II regime boundary.

\subsection{Nesting regime}
\label{subsec:nesting}

\begin{figure}[!ht]
\begin{center}
\vspace*{0.2in}
\includegraphics[width=0.45\textwidth]{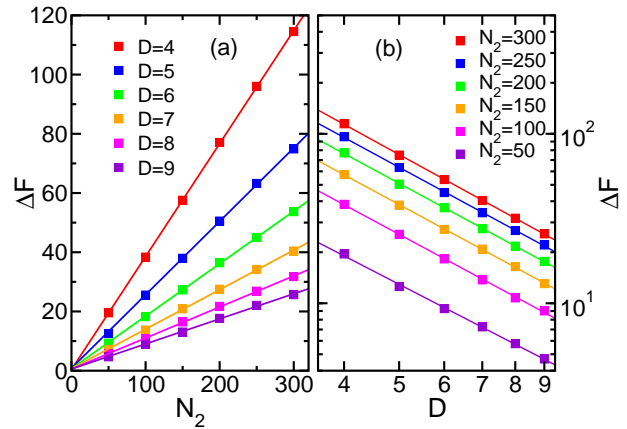}
\end{center}
\caption{(a) Free energy difference $\Delta F$ vs polymer length $N_2$ for two polymers for the
case of $N_1$=200.  Results for various channel diameters are shown.  The solid lines show fits
to a linear function.  (b) $\Delta F\equiv F(\lambda=0)-F(\lambda=\infty)$ vs
channel diameter $D$ for two polymers of length $N_1$=200 and $N_2$ confined to an infinitely
long cylindrical channel. Results are shown for various values of $N_2$. The solid lines
show fits to a power law $\Delta F\sim D^{-\beta}$. The best-fit scaling exponents are:
$\beta=1.84\pm 0.01$ for $N_2$=300, $\beta=1.82\pm 0.02$ for $N_2$=250,
$\beta=1.81\pm 0.01$ for $N_2$=200, $\beta=1.81\pm 0.01$ for $N_2$=150,
$\beta=1.81\pm 0.02$ for $N_2$=100, and $\beta=1.73\pm 0.03$ for $N_2$=50.  }
\label{fig:delF.N1=300}
\end{figure}

Let us now examine regime I in detail. In this regime, the centers of mass are close enough
that the shorter polymer is completely nested within the longer polymer. (Note: an exception
to this rule occurs when the polymer lengths differ by only a small amount, as will be
discussed in Sec.~\ref{subsec:nesting2}.) In addition, the free energy is constant and equal
to the value of the free energy barrier height, which we define $\Delta F\equiv 
F(\lambda=0)-F(\lambda=\infty)$. Since we choose $F(\infty)=0$, it follows that 
$\Delta F = F(\lambda=0)$.  Figure~\ref{fig:delF.N1=300}(a) 
shows the scaling of $\Delta F$ with $N_2$ for fixed $N_1$=300 and for various channel diameters. 
In each case, $\Delta F$ increases linearly with $N_2$, and the slope of the linear functions 
increases with decreasing $D$. Figure~\ref{fig:delF.N1=300}(b) shows the variation of $\Delta F$
with $D$ for fixed $N_1$=300 and for $N_2$=50--300. For arbitrary $N_2$, we find that
$\Delta F \sim D^{-\alpha}$, where $\alpha$ is in the narrow range of $\alpha\approx 1.81-1.84$ 
for all cases except $N_2$=50. Consequently, the $\lambda$-invariant free energy in 
regime~I scales as $F \propto N_2D^{-\alpha}$, where $\alpha\approx 1.82$.

\begin{figure}[!ht]
\begin{center}
\vspace*{0.2in}
\includegraphics[width=0.4\textwidth]{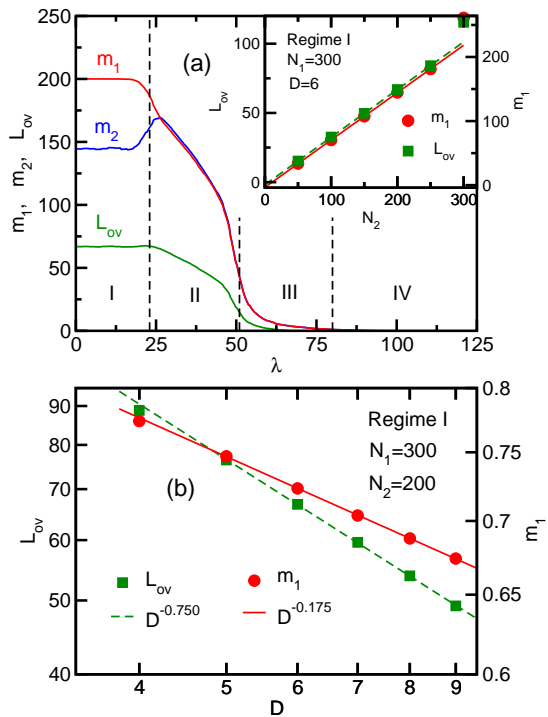}
\end{center}
\caption{(a) Variation of $m_1$, $m_2$, and $L_{\rm ov}$ with $\lambda$ for a system with 
$N_1$=300, $N_2$=200, and $D$=6. As noted in the text, $m_1$ and $m_2$ are the mean number 
of monomers of each polymer that lie in the overlap region. The dashed lines denote the
approximate boundaries of the four overlap regimes. The inset shows the variation of 
$m_1$ and $L_{\rm ov}$ with $N_2$ for fixed $N_1$=300 and $D$=6. (b) Variation of
$L_{\rm ov}$ and $m_1$ with channel diameter for the case where $N_1$=300, $N_2$=200.
The solid lines are fits to power laws.  }
\label{fig:regime_I_scale}
\end{figure}

Let $m_1$ and $m_2$ be the average number of monomers that lie in the overlap range for
any separation $\lambda$.  Figure~\ref{fig:regime_I_scale}(a) shows variation of $m_1$
and $m_2$ with $\lambda$ for a case where $N_1$=300, $N_2$=200, and $D$=6. In the case
of partial overlap in regime~II, we see that $m_1\approx m_2$ and that both increase
as the separation decreases. Upon further decreasing $\lambda$, the system enters
regime~I, and the values of $m_1$ and $m_2$ diverge. Since the shorter polymer is now
completely nested inside the longer polymer, it follows that $m_2=N_2=200$.
On the other hand, number of monomers from the longer polymer in the overlap region
abruptly decreases to a constant value that satisfies $m_1<m_2$. The inset of 
Figure~\ref{fig:regime_I_scale}(a) shows that both the value of $m_1$ as well as the 
overlap distance $L_{\rm ov}$ both increase linearly with increasing $N_2$. The
exception is the limiting case where $N_2=N_1=300$. 

Figure~\ref{fig:regime_I_scale}(b) shows the variation of $L_{\rm ov}$ and $m_1$ with 
$D$ in regime~I for the case where $N_1$=300 and $N_2$=200. In both cases, power-law
scaling is observed. The overlap distance scales as $L_{\rm ov}\sim D^{-0.75}$, while
$m_1$ has a much weaker dependence on $D$, with a scaling of $m_1\sim D^{-0.175}$.

To understand the scaling behavior of Figs.~\ref{fig:delF.N1=300} and \ref{fig:regime_I_scale}
 we use a simple theoretical model that is based on two approximations. First, the scaling behavior 
of the free energy is assumed to correspond to the predictions of the blob model in the de~Gennes 
confinement regime.\cite{deGennes_book} A second approximation is that introduced by
Jung {\it et al.} to approximate the free energy of overlapping confined chains.\cite{jung2012ring}
In this picture, the effect of longitudinal overlap on the conformational behavior of the two 
channel-confined polymers is approximately the same as that of confining the overlapping portions 
of the chains to virtual tubes of reduced cross-sectional areas. The overlapping portion of 
polymer \#1 is pictured as confined to a virtual tube of diameter $D_1=D\sqrt{\xi}$, where $\xi<1$, 
while the overlapping portion of polymer \#2 is likewise confined to a virtual tube of diameter 
$D_2=D\sqrt{\xi_0-\xi}$.  Jung {\it et al.} used this model to describe the conformational behavior of
of a ring polymer confined to a channel. Since the ring polymer was modeled as two completely overlapping 
linear polymers of equal length, $D_1=D_2$. Neglecting lateral interpenetration requires the cross-sectional 
areas of the virtual tubes sum to equal that of the actual channel. Together, these assumptions imply
that $\xi=\frac{1}{2}$ and $\xi_0=1$. Choosing $\xi_0 \geq 1$ effectively allows for some lateral
interpenetration, while $\xi_0-\xi \neq \xi$ (i.e. $D_1\neq D_2$) allows for unequal virtual
tube sizes, which is clearly required for regime~I, where $m_1\neq m_2$. Note that $\xi$ and $\xi_0$ 
are assumed to be independent of $D$.

In Appendix~\ref{app:a} we show that this theoretical model predicts that the free energy barrier 
height scales as 
\begin{eqnarray}
\Delta F \propto N_2 D^{-1/\nu} f(m_1/N_2;\xi_0),
\label{eq:Fpropto}
\end{eqnarray}
 where $f(m_1/N_2;\xi_0)$ is 
defined in Eq.~(\ref{eq:fxbeta}). Thus, the model correctly predicts the linear scaling of 
$\Delta F$ with respect to $N_2$ observed in Fig.~\ref{fig:delF.N1=300}(a). The model also
predicts that $\Delta F\sim D^{-1.70}$, where we use the Flory exponent value $\nu\approx 0.588$.
Thus, the predicted scaling exponent is close to measured exponent of $\approx 1.82$ obtained from 
fits to the data in Fig.~\ref{fig:delF.N1=300}(b). A similar discrepancy was observed for equal-length
linear and ring polymers in our previous study.\cite{polson2018segregation} 

The origin of the discrepancy between the values of the measured and predicted scaling exponent is 
likely a combination of two factors. The first is a breakdown in the blob-model scaling behavior 
for the narrow channel widths and polymer lengths considered here. As noted in 
Ref.~\onlinecite{kim2013elasticity} such scaling requires that $D$ and $N$ be sufficiently 
large that both the number of monomers per blob, $g\sim D^{1/\nu}$, and the number of blobs, 
$n_{\rm blob}=N/g$, be large compared to unity. A second factor is a breakdown in the validity
of the approximation used to estimate the overlap free energy function in Eq.~(\ref{eq:Fpropto}). 
The accuracy of this approximation was measured directly in Ref.~\onlinecite{polson2018segregation}, 
and the simulations revealed quantitative inaccuracies arising from finite-size effects. 
Additional insight into this discrepancy is provided by Eq.~(\ref{eq:xalpha}) in Appendix~\ref{app:a}, 
where the theory predicts that $m_1$ should be independent of the channel width. This contradicts
the results in Fig.~\ref{fig:regime_I_scale}(b), which shows a weak but non-negligible dependence 
of $m_1$ on $D$. In addition, the overlap length is predicted to scale as 
$L_{\rm ov} = X_2\propto N_2 D_2^{(\nu-1)/\nu} = N_2 (D\sqrt{\xi_0-\xi})^{(\nu-1)/\nu} \propto D^{-0.70}$.
The predicted scaling exponent is close to the value measured from the data in Fig.~\ref{fig:regime_I_scale}(b),
though once again the discrepancy is non-negligible. 

We suspect that the difference between the predicted and measured scaling exponents arises from the 
assumption that the factors $\xi$ and $\xi_0$ are independent of $D$. While incorporation of such a 
dependence could improve the theory, there are no obvious physical arguments to suggest any particular 
functional form. Despite these limitations, the reasonably accurate predictions suggest that the theoretical 
model provides a decent understanding of the overlap free energy scaling behavior.  
Calculations presented in Ref.~\onlinecite{polson2018segregation} suggest that using constant values
of the quantities $\xi$ and $\xi_0$ may be justified, but only in the limit of much larger $N$ and 
$D$. As noted in that study, no single value of the rescaling factor yields predictions of high 
quality for the polymer lengths and channel width employed here.\cite{polson2018segregation}

\subsection{Nesting free energy function for $\lambda=0$}
\label{subsec:nesting2}

Let us consider further the system behavior in the nesting regime in the case of overlapping
centers of mass, i.e., $\lambda$=0.  This state was examined in detail by Minina and 
Arnold\cite{minina2014induction,minina2015entropic}
for the case of two polymers of equal length. They noted that the separation kinetics of
polymers that begin at $\lambda=0$ show an initial lag time, which they attributed to
a free energy barrier associated with the nesting of one polymer within the other. 
We examined the scaling behavior of this free energy barrier previously for two identical
polymers.\cite{polson2018segregation} Here, we extend that study and examine the effects
of small differences in the lengths of the polymers on the barrier. 

\begin{figure}[!ht]
\begin{center}
\vspace*{0.2in}
\includegraphics[width=0.4\textwidth]{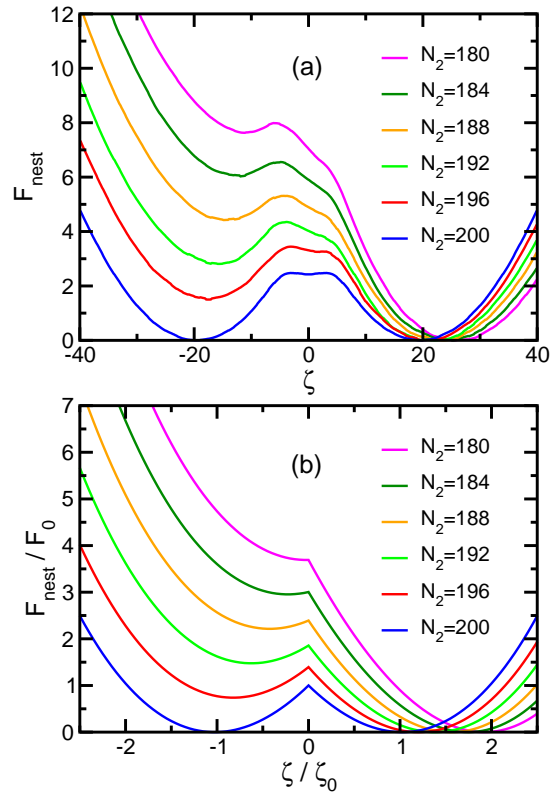}
\end{center}
\caption{(a) Free energy vs extension length difference $\zeta\equiv X_1-X_2$ for polymers
with overlapping centers of mass. Results are shown for $D$=4, $N_1$=200 and for various
values of $N_2$. (b) Prediction of the nesting free energy using the theory developed in
Appendix~\ref{app:a}. The scaling factors for the axes are defined with respect to the
curve for $N_1=N_2=200$: (i) $F_0\equiv F_{\rm nest}(0)$; and (ii) $\zeta_0$ is the
value of $\zeta$ at the right minimum of $F_{\rm nest}$.}
\label{fig:nesting}
\end{figure}

Following Ref.~\onlinecite{polson2018segregation} we calculate the free energy at $\lambda=0$ 
as a function of $\zeta \equiv X_1-X_2$, i.e. the difference in the extension lengths of the polymers. 
When $N_1=N_2$, it is equally likely that either polymer is nested inside the other.
Consequently, two identical free energy minima are observed separated by a small free energy 
barrier.\cite{polson2018segregation,minina2014induction,minina2015entropic}
Such a symmetric free energy function is shown in Fig.~\ref{fig:nesting}(a) in the case
of $N_1=N_2=200$ for $D=4$.  As $N_2$ is decreased
to values below $N_1$, the double-well structure is initially preserved, though the free energy
at the minimum at lower $\zeta$ tends to increase relative to the minimum at higher $\zeta$.
Thus, provided $N_2$ is not too much less than $N_1$, the longer polymer nested inside the
shorter polymer represents a metastable state separated from the global free energy minimum
(where the roles of the polymers are reversed) by a free energy barrier. The height of the
barrier decreases with decreasing $N_2$ until it almost disappears at $N_2$=180.

Some understanding of the origins of the trends observed in Fig.~\ref{fig:nesting}(a) is
provided by the theoretical model developed by Minina and Arnold.\cite{minina2014induction,%
minina2015entropic} They employed essentially the same two approximations as were used to elucidate
the scaling behavior observed in Fig.~\ref{fig:delF.N1=300} above: (i) using the de~Gennes
blob model to describe the scaling of the free energy and extension lengths of channel-confined
polymers; and (ii) assuming that the effect of overlap on the free energy is the same as
that of separately confining each polymer to a different virtual tube, the sum of whose
cross-sectional areas is equal to that of the real channel.\cite{jung2012ring} In Appendix~\ref{app:b}
we modify their theory to accommodate polymers of unequal length. The theoretical predictions
are shown in Fig.~\ref{fig:nesting}(b). As the theory employs scaling relations with
undetermined prefactors, the axes have been scaled by factors as defined in the caption to
avoid confusion.
Although the shape of the free energy barrier is poorly described by the theory, the key
qualitative behavior is well reproduced. In particular, as the length asymmetry increases
(i.e. $N_2$ decreases while $N_1$ is held fixed), the position of the free energy minima both
shift to higher $\zeta$, and the free energy of the negative-$\zeta$ minimum increases relative
to that of the positive-$\zeta$ minimum. In addition, the metastable state eventually disappears
as $N_2$ decreases. As before, quantitative discrepancies arise from the nature of the 
approximations employed.

\subsection{Scaling behavior of $F(\lambda)$}
\label{subsec:scaling}

\begin{figure}[!ht]
\begin{center}
\vspace*{0.2in}
\includegraphics[width=0.45\textwidth]{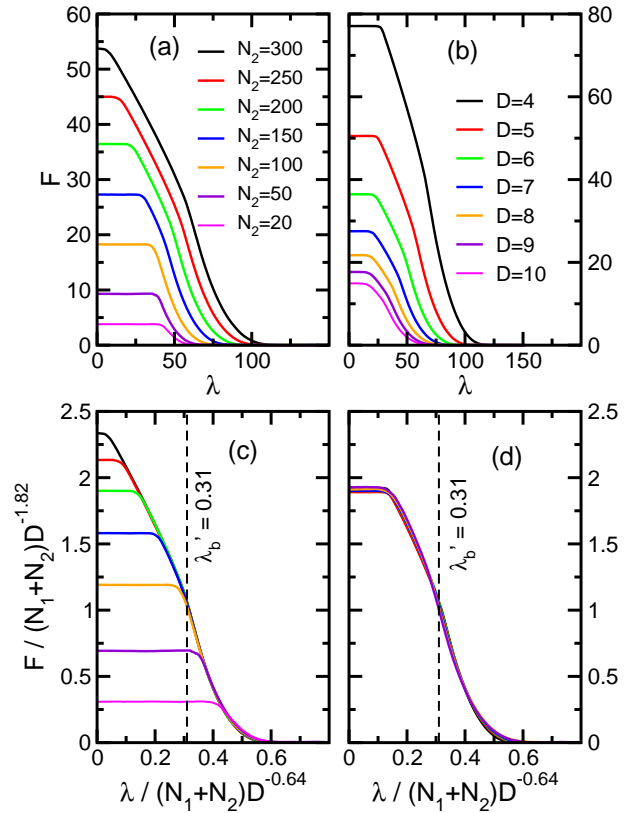}
\end{center}
\caption{(a) Free energy vs $\lambda$ for polymers confined to a cylinder of diameter $D$=6. 
One polymer length is fixed to $N_1$=200. Results for various values of the polymer length $N_2$ 
($\leq N_1$) are shown. (b) $F$ vs $\lambda$ for polymers of length $N_1$=300 and $N_2$=200 for 
various cylinder diameters. (c) $F^\prime\equiv F/(N_1+N_2)D^{-1.82}$ vs $\lambda^\prime\equiv 
\lambda/(N_1+N_2)D^{-0.64}$ using the data in panel (a). (d) As in (c) except using the data in 
panel (b). The vertical dashed lines in panels (c) and (d) mark the approximate boundary at 
$\lambda^\prime_{\rm b}\approx 0.31$ between regimes II and III for $N_2\gtrsim 100$. }
\label{fig:F.N1=300}
\end{figure}

Thus far we have focused on the behavior of the system solely in regime~I. We now examine 
the scaling properties of $F(\lambda)$ over the full range of $\lambda$ across all regimes.
Figure~\ref{fig:F.N1=300}(a) shows free energy functions for $N_1$=300 and $D$=6 and for 
various values of the length $N_2$. The linear increase of $\Delta F$ with $N_2$ has already 
been discussed at length in Section~\ref{subsec:nesting}. The increasing width of the free energy 
barrier with increasing $N_2$ can be understood as follows. In the case of isolated 
polymers with large $\lambda$, the extension lengths are each proportional to the contour 
length. Thus, as the polymers are brought closer together, the center-of-mass distance $\lambda$ 
corresponding to the point where the polymers make contact (i.e. at the boundary of regime III 
and IV) is shorter for smaller $N_2$. Another trend is the decreasing width of regime~I as 
$N_2$ increases and approaches $N_1$. To understand this trend, we note that the transition to regime~
II occurs at the value of $\lambda$ where the location of the nested polymer's left (right) boundary
along the channel reaches the left (right) boundary of the longer polymer. This occurs when 
$\lambda=(X_1-X_2)/2 = (X_1-c_2 N_2)/2$, since $X_2\propto N_2$ in the nesting regime
and where $c_2$ is the proportionality constant.  Thus, the value of $\lambda$ at this transition 
decreases with increasing $N_2$, thereby reducing the width of regime~I. Finally, the boundary 
between regimes~II and III evident from the inflection point in the functions disappears
for $N_2\lesssim 100$.

Figure~\ref{fig:F.N1=300}(b) shows free energy functions for $N_1$=300 and $N_1$=200 for various 
values of the cylinder diameter $D$. As $D$ increases, we note that the barrier height
and width both decrease. The latter trend is due to the fact that reducing the confining tube 
width also reduces the extension of the polymers. Consequently, the polymer center-of-mass 
separation distance at the II-III and III-IV regime boundaries is smaller for larger $D$.

The quantitative nature of the scaling of $F(\lambda)$ with respect to $N_2$ and $D$ can be
elucidated by considering the scaled variables $F^\prime \equiv F/(N_1+N_2)D^{-\alpha}$ and 
$\lambda^\prime \equiv \lambda/(N_1+N_2)D^{-\beta}$. Figures~\ref{fig:F.N1=300}(c) and 
\ref{fig:F.N1=300}(d) show the results such a transformation on the free energy functions 
of Figs.~\ref{fig:F.N1=300}(a) and \ref{fig:F.N1=300}(b), respectively, using
exponents of $\alpha=1.82$ and $\beta=0.64$.  The results are revealing. 
In Fig.~\ref{fig:F.N1=300}(c), all of the functions for the various values of $N_2$ collapse 
to a universal curve, except at low $\lambda^\prime$, where the system is in regime~I. 
The universal curve can be defined as the function for $N_1=N_2$. Define $\lambda^\prime_{\rm b}$
to be the value of $\lambda^\prime$ at the boundary between regimes II and III for this curve.
Clearly, the boundary is at $\lambda^\prime_{\rm b}\approx 0.31$, and is marked on the graph with a 
dashed line. Now define $\lambda^\prime_*$ as the value of $\lambda^\prime$ 
at which the regime-I plateau meets the universal curve. We see that $\lambda^\prime_*$ 
increases as the polymer length $N_2$ decreases. In the case where $N_2\gtrsim 100$, 
$\lambda^\prime_* < \lambda^\prime_{\rm b}$, which simply implies that regime II exists for this
range of $N_2$. On the other hand, for $N_2\lesssim 100$, $\lambda^\prime_* > \lambda^\prime_{\rm b}$.
This means that regime~II is absent for these shorter polymer lengths, and thus the system
passes directly from the nesting regime (I) to the compressed regime (III).
In Fig.~\ref{fig:F.N1=300}(d), all of the scaled free energy functions for different $D$
and fixed $N_2$ collapse to a universal curve in all regimes, including regime~I. 

The results above suggest the following scaling for the free energy for regimes II, III
and IV:
\begin{eqnarray}
F(\lambda)=(N_1+N_2)D^{-\alpha} f(\lambda/(N_1+N_2)D^{-\beta}), 
\label{eq:Fscale}
\end{eqnarray}
where $f()$ is a universal function and
where $\alpha\approx 0.64$ and $\beta\approx 1.82$. (Since $F=0$ in regime IV, the scaling 
is automatically satisfied here.) This is consistent with the case of symmetric systems with
$N_1=N_2\equiv N$, for which we previously showed that $F(\lambda)=ND^{-\alpha} f(\lambda/ND^\beta)$
with approximately equal exponent values.\cite{polson2018segregation} In the case of regime~I, we have 
already noted in Section~\ref{subsec:nesting}  that $F$ is independent of $\lambda$ and scales as 
\begin{eqnarray}
F \propto N_2 D^{-\alpha}.
\label{eq:FI}
\end{eqnarray}
For $N_1$=300 and $N_2$=200 the fit to the data in Fig.~\ref{fig:delF.N1=300}(b) 
yielded $\alpha\approx$1.84, which is very close to the value of $\alpha$=1.82 that
best collapses the curves in Fig.~\ref{fig:F.N1=300}(d). Although the scaling relations
were obtained from analysis of data for $N_1$=300, the results do not change for other
values of $N_1$. In Appendix~\ref{app:c}, this is illustrated by a
comparison of free energy functions calculated for $N_1$=200, 300, and 400.
The I-II and I-III regime boundary is determined by the condition the free energy in
Eqs.~(\ref{eq:Fscale}) and (\ref{eq:FI}) are equal. This leads to the condition that
$f(\lambda^\prime) \propto (1+(N_2/N_1)^{-1})^{-1}$, where $\lambda^\prime\equiv 
\lambda/(N_1+N_2)D^{-\beta}$ and where the proportionality constant is the same as 
in Eq.~(\ref{eq:FI}). Thus, the I-II and I-III regime boundaries are uniquely determined by 
the quantities $\lambda^\prime$ and $N_2/N_1$, as implicitly true in Eq.~(\ref{eq:Fscale})
for the II-III and III-IV boundaries. 

\begin{figure}[!ht]
\begin{center}
\vspace*{0.2in}
\includegraphics[width=0.4\textwidth]{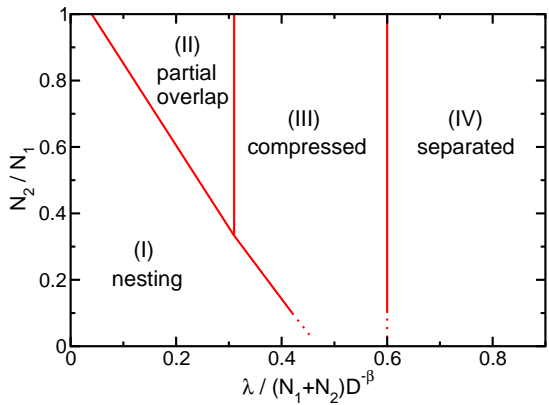}
\end{center}
\caption{Proposed regime map for two polymers of length $N_1$ and $N_2$ ($\leq N_1$) separated
by center-of-mass distance $\lambda$ confined to a cylindrical channel of diameter $D$.
The II-III and III-IV phase boundaries are vertical lines, as shown. However, the I-II and I-III
boundaries are only qualitatively illustrated. Simulations with $N_1$=300 yield an exponent of
$\beta\approx$0.64. In regimes II, III and IV, the free energy functions scale according to
Eq.~(\ref{eq:Fscale}), and in regime~I the free energy satisfies Eq.~(\ref{eq:FI}). }
\label{fig:phase}
\end{figure}

Figure~\ref{fig:phase} shows a ``phase diagram'' illustrating the approximate boundaries
separating the four scaling regimes. Our previous study\cite{polson2018segregation} covered 
the special case of $N_2/N_1=1$, which corresponds to a horizontal line at the very top
of this diagram. Clearly, the present study extends the characterization of the system to 
cover a much larger region of parameter space. Of special note is the disappearance of the
partial overlapping regime for $N_2/N_1\lesssim 0.3$, i.e., for a sufficiently large
polymer size asymmetry. This behavior has a simple analogue in the phase behavior of
substances like CO$_{\rm 2}$ or water, i.e., the disappearance of a liquid phase upon
heating a solid at a fixed pressure below that of the triple-point pressure so that the solid
sublimes directly to a gas.

In order to understand the regime map of the system described above, we follow an approach
similar to that in our previous study.\cite{polson2018segregation} Since we have already
explained the scaling of regime~I in Section~\ref{subsec:nesting}, and since the scaling
of $F=0$ in regime~IV follows Eq.~(\ref{eq:Fscale}) trivially, we need only deal with
regimes~II and III. We begin with the latter.  Recall that regime~III corresponds to the case 
where the polymers are compressed and in contact, but not overlapping. Here, the two polymers
behave much like a single channel-confined polymer in a compressed state.  Employing the 
renormalized Flory theory of Jun {\it et al.},\cite{jun2008compression} the free energy for a 
single linear polymer of length $L_{\rm ext}$ is
\begin{eqnarray}
F=AL_{\rm ext}^2/(N/g)D^2 + BD(N/g)^2/L_{\rm ext},
\label{eq:Frenorm}
\end{eqnarray}
where $A$ and $B$ are constants of order unity and $g\sim D^{1/\nu}$ is the number of
monomers in a compression blob of diameter $D$. Noting $\lambda=L_{\rm ext}/2$
(i.e. assuming uniform compression) it is easily shown that
\begin{eqnarray}
F(\lambda) = (N_1+N_2) D^{-1/\nu} w(\lambda/(N_1+N_2)D^{1-1/\nu}),
\label{eq:FlinIII}
\end{eqnarray}
where $w(x) = 4Ax^2+B/2x$. This is consistent in form with the scaling of Eq.~(\ref{eq:Fscale}),
with scaling exponents of $\alpha=1/\nu$ and $\beta=1/\nu-1$. Using $\nu=0.588$, this
corresponds to a prediction of $\alpha=1.70$ and $\beta=0.70$, which are close to the
measured values of $\alpha=1.82$ and $\beta=0.64$.

To understand the scaling of $F(\lambda)$ in regime~II, we use the same theoretical
models developed in Appendices~\ref{app:a} and \ref{app:b} to understand the scaling of the
free energy barrier height. Specifically, we employ the blob scaling model together with 
the approximation introduced by Jung~{\it et al.}\cite{jung2012ring} to account for the
conformational behavior of overlapping chains in channels. The calculation is presented
in Appendix~\ref{app:d}.  In Eq.~(\ref{eq:FN1N2D}), we find that $F/(N_1+N_2)D^{-1/\nu} 
\propto m/(N_1+N_2)$, where $m\equiv m_1=m_2$ is the number of overlapping monomers per polymer. 
In order for Eq.~(\ref{eq:Fscale}) to be satisfied in regime~II, we expect $m/(N_1+N_2)$ to 
be a universal function of $\lambda/(N_1+N_2)D^{1-1/\nu}$ independent of the values of $N_1$ 
and $N_2$. As described in the appendix, the observed variation of $m/(N_1+N_2)$ with 
$\lambda/(N_1+N_2)D^{1-1/\nu}$ does show some slight dependence on $N_2$ and $N_1$ (though
not on $D$), but this appears to be negligible over the range of $\lambda$ corresponding
to regime~II. As for regime~III the scaling exponents for Eq.~(\ref{eq:Fscale})
are predicted to be $\alpha = 1/\nu\approx 1.70$ and $\beta=1/\nu-1\approx 0.70$. Again,
the deviations between the predicted and measured exponents arise from the nature of the
approximations used.

\subsection{Segregation Dynamics}
\label{subsec:dynamics}

We now consider the dynamics of polymers segregating from an initial state of maximum 
overlap at $\lambda$=0. Several other simulation studies have investigated the segregation
dynamics of comparable systems. In the present work, our goal is to elucidate the relationship,
if any, between the dynamics and the equilibrium free energy functions. This extends
the range of calculations carried out in our previous study,\cite{polson2014polymer} which
considered only the case of polymers of equal length. 

\begin{figure}[!ht]
\begin{center}
\vspace*{0.2in}
\includegraphics[width=0.45\textwidth]{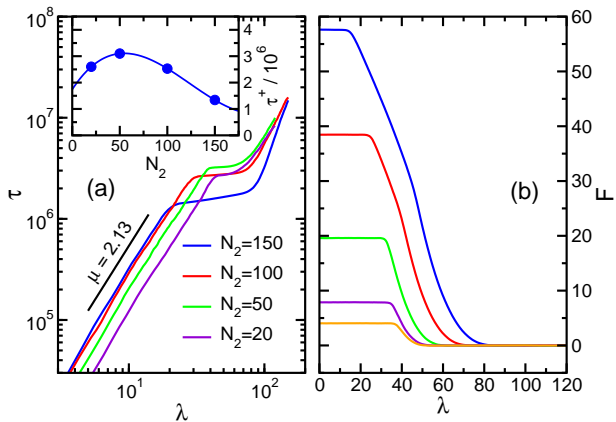}
\end{center}\caption{(a) Incremental mean-first-passage time $\tau$ vs $\lambda$ for segregating
polymers. Results are shown for $D$=4, $N_1$=200 and for various values of $N_2$.
The inset shows the variation of $\tau^\dag$ with $\lambda$, where $\tau^\dag$ is
the mean first-passage time for the system to enter the middle dynamical regime,
as explained in the text.  (b) Free energy functions for the systems described in (a).  }
\label{fig:imfpt_N1=200_R=2.5}
\end{figure}

Figure~\ref{fig:imfpt_N1=200_R=2.5}(a) shows the variation of the 
IMFPT with $\lambda$ for the case of segregating polymers.
Results are shown for a confinement tube diameter of $D$=4, $N_1$=200 and for various values of 
$N_2$. As a useful reference, Fig.~\ref{fig:imfpt_N1=200_R=2.5}(b) shows the free energy functions 
for each of the systems considered in (a). At low $\lambda$, we find that $\tau$ obeys a power 
law of the form $\tau\sim \lambda^\mu$, where $\mu\approx 2.1$. This dynamical regime 
approximately corresponds to regime~I of the free energy functions, i.e. the regime in which $F$ is
invariant with respect to $\lambda$. In addition, at any separation distance $\lambda$
in this regime $\tau$ increases with increasing $N_2$. A simple explanation for these
trends follows from modeling the polymers as two independent 1D random walkers with diffusion
coefficients ${\cal D}_1$ and ${\cal D}_2$. In this case, the distance between the two walkers 
satisfies a diffusion equation with effective diffusion coefficient 
${\cal D}_{\rm eff}={\cal D}_1+{\cal D}_2$. Starting from a separation $\lambda_0=0$, the 
first-passage time required to reach a separation of $\pm\lambda$ for this simple diffusion 
is $\tau = \lambda^2/(24({\cal D}_1+{\cal D}_2))$.\cite{redner_book} Thus, the non-interacting-walkers model 
predicts an exponent of $\mu=2$, which is close to the observed value. Since ${\cal D}_2\propto 1/N_2$, 
it follows that $\tau$ will increase as the $N_2$ increases, consistent with the results.

The variation of $\tau$ with separation distance changes abruptly at a $N_2$-dependent
value of $\lambda$ after which it increases at a much slower rate. Following this stage,
there is a second transition at a higher value of $\lambda$, after which $\tau$
increases rapidly once more with separation distance. The middle regime corresponds roughly
to regimes~II and III of the free energy functions. Thus, it corresponds to the region
in which the short polymer has left the nested stage and is either partially overlapping 
with the long polymer or else is not overlapping but is in contact and compressed. 
Significantly, these regimes correspond to the rapid decrease in the free energy. The
gradient of $F$ can be considered an effective force that drives the polymers apart.
While the system is subject to this effective force, any additional increase in the 
separation distance is achieved in much less time than is the case in its absence, leading
to the observed behavior. Once the polymers are no longer in contact (i.e. regime~IV) 
the separation dynamics are once again governed by the much slower process of unbiased 
diffusion, resulting in a more rapid increase in $\tau$ with $\lambda$. 

Another interesting trend is the variation of $\tau$ with the length of the short polymer, 
$N_2$, in the middle dynamical regime.  Unlike the case at low $\lambda$ where
$\tau$ increases monotonically with $N_2$, this is clearly not the case here. To
clarify this trend, we define $\tau^\dag$ as the mean first-passage time for when
the polymers reach a separation distance corresponding to the onset of the middle
dynamical regime. The variation of $\tau^\dag$ with $N_2$ is shown in the inset of
Fig.~\ref{fig:imfpt_N1=200_R=2.5}(a). The variation is non-monotonic and has a maximum
$\tau^\dag$ at $N_2\approx 60$. The origin of this behavior stems from two competing trends. 
First, we note that $\tau^\dag$ is approximately the time required for the system to reach 
the boundary between regimes~I and II shown in Fig.~\ref{fig:F_illust}. As noted in Sec.~\ref{subsec:scaling}
this occurs at $\lambda\approx (X_1-X_2)/2= (X_1-c_2N_2)/2$. Thus, the separation distance 
at this boundary increases as $N_2$ decreases. In turn, this will tend to cause the time
$\tau^\dag$ to increase. On the other hand, decreasing $N_2$ also increases ${\cal D}_2$ and 
therefore also the relative diffusion rate ${\cal D}_1+{\cal D}_2$. This effect will tend 
to decrease $\tau^\dag$.  Evidently, at high $N_2$ the first effect dominates, while the 
second effect dominates at low $N_2$, thus resulting in the maximum.

\begin{figure}[!ht]
\begin{center}
\vspace*{0.2in}
\includegraphics[width=0.4\textwidth]{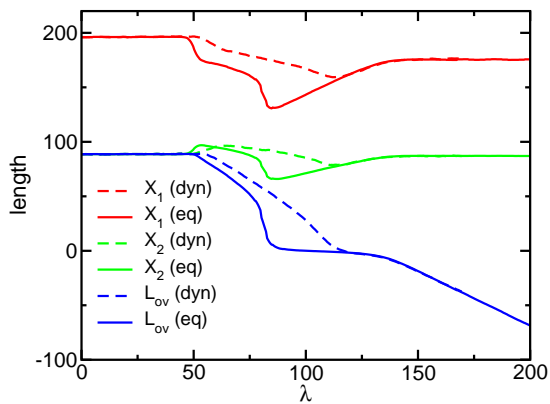}
\end{center}
\caption{Extension lengths and overlap distance vs $\lambda$. The graph compares the results
obtained from equilibrium simulations with those obtained from a MC dynamics simulation
for segregating polymers that initially overlap such that $\lambda$=0. Results are shown
for $N_1$=400, $N_2$=200 and $D$=4.}
\label{fig:Ldyn}
\end{figure}

As noted above, the two transitions in the segregation kinetics occur near the I-II and
III-IV regime boundaries of the corresponding free energy functions. A close inspection
reveals that the first transition in the kinetics occurs at a separation distance that is
consistently larger than the I-II regime boundary. For example, for $N_2$=150, the 
transition occurs at $\lambda\approx 20$, while the I-II regime boundary is located at
$\lambda\approx 14$. To understand this apparent discrepancy it is useful to examine once
again the variation of the extension lengths and overlap distance with $\lambda$. In 
particular, we compare the behavior of these functions calculated during the segregation
simulations with those calculated during the (equilibrium) calculations of the free energy 
functions.  Figure~\ref{fig:Ldyn} shows the two sets of results calculated for a system 
with $N_1$=400, $N_2$=200 and $D$=4. In regime~I, the curves for each extension length 
calculated using the two methods overlap, and likewise for $L_{\rm ov}$.
This is also true in regime~IV, where the polymers are no longer in physical contact. However,
there is a significant difference between the two sets of results in the intermediate regimes.
This indicates that segregation is not a quasi-static process in this segregation stage, which
explains the small discrepancies between the results of Figs.~\ref{fig:imfpt_N1=200_R=2.5}(a)
and (b). For example, the transition associated with leaving the nesting state occurs at
slightly a higher $\lambda$ value in the results from the dynamics simulations relative to 
the equilibrium case, consistent with trends discussed above in relation to the behavior 
of the  IMFPT.  

Another noteworthy difference is the absence of any clear transition
in the dynamics results associated with the regime II-III boundary of the free energy function.
The reason that the two sets of results diverge in this region is straightforward: it 
corresponds roughly to the regime where the driving force is greatest. Thus, the polymers
are driven apart at a rate that is fast relative to that of the internal relaxation and
out-of-equilibrium effects become significant. Once the driving force reduces to zero
and unbiased diffusion dominates, then quasi-equilibrium is restored.

We conclude with the following two observations concerning the relation between the
segregation dynamics and the free energy functions. First, the separation kinetics in the
nesting regime, where the system is in quasi-equilibrium, exhibits behavior consistent
with the flat free energy function for that regime. Second, a rapid increase in the 
separation rate occurs at $\lambda$ very close to values corresponding to steep gradients
in the free energy. Since out-of-equilibrium effects are present during this stage, 
the equilibrium free energy cannot be used in any quantitative analysis of the dynamics.
However, the free energies do provide information on where such a rapid increase in
the segregation rate is expected to occur.

\section{Conclusions}
\label{sec:conclusions}

In this study we have used MC simulations to study the physical behavior of two flexible
hard-sphere polymer chains confined to infinitely-long channels. Building on our previous
studies of symmetric polymer/channel systems,\cite{polson2014polymer,polson2018segregation} 
we consider the effects of asymmetry in the polymer length. The main focus was 
the measurement and characterization of the variation of the free energy $F$ with respect to 
the center-of-mass separation along the channel, $\lambda$. For the case where the polymer 
centers of mass are constrained to overlap, we also measure $F$ with respect to extension 
length difference $\zeta$. In addition, we used MC dynamics simulations to study the segregation 
dynamics of initially overlapping polymers.

A summary of the key findings is as follows.
The free energy functions $F(\lambda)$ exhibited the same four regimes ((I) nested, (II) partial 
overlap, (III) compressed/non-overlap, and (IV) separated) as observed in our previous studies 
on polymer segregation.\cite{polson2014polymer,polson2018segregation} In Fig.~\ref{fig:phase} 
we propose a regime map in which the state of the system is determined by the ratio of
the polymer lengths, $N_2/N_1$, and the scaled separation distance, 
$\lambda^\prime\equiv \lambda/(N_1+N_2)D^{-\beta}$.  For sufficiently small $N_2/N_1$, regime~II
disappears and the system passes directly from regime~I to III as $\lambda^\prime$ increases.
In regimes~II, III and IV, the free energy functions satisfy the scaling
$F(\lambda) = (N_1+N_2)D^{-\alpha} f(\lambda/(N_1+N_2)D^{-\beta})$, where $\alpha\approx 1.82$.
and $\beta\approx 0.64$. In regime~I, where the short polymer is nested inside the larger one, $F$ 
is constant with respect to $\lambda$ and satisfies $F\propto N_2D^{-\beta}$, where $\beta\approx 0.64$. 
A simple theory constructed using de~Gennes blob scaling and a previously proposed model to describe 
the conformational behavior of overlapping polymers in channels\cite{jung2012ring} yields scaling 
behavior consistent with the observed behavior and predicts exponents of $\alpha=1/\nu\approx 1.70$
and $\beta=1-1/\nu\approx 0.70$. The small discrepancies arise from a combination of finite-size
effects and deficiencies in the theoretical model of Ref.~\onlinecite{jung2012ring}.
Regime~I is also characterized by dynamics that are surprisingly similar to those expected 
for two noninteracting 1D random walkers undergoing unbiased diffusion. When the system crosses 
into regimes II and III, a rapid segregation occurs with out-of-equilibrium conformational behavior.

Our study has revealed a few intriguing effects in systems of polymers confined to channels 
related to the asymmetry in the contour lengths. The most notable are associated with the 
proposed ``phase diagram'' sketched in Fig.~\ref{fig:phase}. In future work, it would be 
of interest to verify that the measured scaling of the free energy and the regime boundaries 
are preserved for much larger polymer systems and to develop theoretical models that yield
more accurate predictions for the scaling exponents. In the future, we also plan to study
the effects of other asymmetries, such as differences in polymer topology (e.g. ring and
linear polymers), as well as asymmetries in the two lateral dimensions of the confining
channel (e.g. using channels with rectangular cross sections).

It must be noted that this study and others like it are still very far removed 
from the long-term goal of accurately quantifying the degree to which entropic forces drive
chromosome segregation in bacteria. Given the inherent complexity of such a biological system,
the models are likely too simplistic for that ambitious goal, even with further refinements.
A more realistic direction is modeling the behavior of ``cleaner'' {\it in vitro} systems
such as DNA and other polymers confined to nanofabricated channels and cavities, as examined 
recently for example in Ref.~\onlinecite{capaldi2018probing}. We anticipate further such
experimental examinations of systems of multiple polymers under confinement, the relevance of
which is aligned more with development of nanofluidic devices for DNA manipulation and analysis. 
Accurate modeling of even these ``simpler'' systems is still challenging. At the very least, 
this requires using much longer, semiflexible chains that better describe polymers such as 
$\lambda$-DNA used in experiments. The type of asymmetry examined in this study may well be 
relevant to such future experiments.

\begin{acknowledgments}
This work was supported by the Natural Sciences and Engineering Research Council of Canada (NSERC).  
We are grateful to Compute Canada and the Atlantic Computational Excellence Network (ACEnet) for 
use of their computational resources.
\end{acknowledgments}

\appendix

\section{Free energy of overlapping polymers in regime~I}
\label{app:a}

In this appendix, we develop a theoretical model for the free energy barrier height, $\Delta F$, 
for channel-confined polymers of unequal chain length. The model is used to help understand the
scaling properties of $\Delta F$ presented in Sec.~\ref{subsec:nesting}.  We follow the approach
taken by Minina and Arnold in Refs.~\onlinecite{minina2014induction} and \onlinecite{minina2015entropic}, 
which considered the special case of equal chain length.

\begin{figure}[!ht]
\begin{center}
%\vspace*{0.2in}
%\includegraphics[width=0.4\textwidth]{figures/nestblob.eps}
\includegraphics[width=0.4\textwidth]{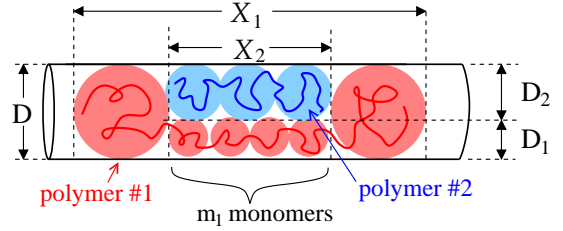}
\end{center}\caption{Illustration of the various quantities described in the text of
Appendix~\ref{app:a} for the case of $\zeta\equiv X_1-X_2 > 0$. For the other case,
where $\zeta\equiv X_1-X_2 < 0$, the polymer \#1 (colored red) would be nested inside
of polymer \#2 (colored blue).}
\label{fig:nestblob}
\end{figure}

Consider two polymers confined to a channel of diameter $D$ in the case where 
polymer \#2 of length $N_2$ is nested in a polymer \#1 of length $N_1$. Thus,
the extension length difference satisfies $\zeta \equiv X_1-X_2 > 0$. Employing the model
of Jung {\it et al.} to describe overlapping polymers under confinement,\cite{jung2012ring}
the overlapping portions of the polymers effectively occupy separate virtual tubes of 
diameter $D_1$ and $D_2$, both of which are less than $D$. 
We write $D_1=D\sqrt{\xi}$ and $D_2=D\sqrt{\xi_0-\xi}$.
In the case where lateral interpenetration is excluded, the sum of the cross sectional
areas of the virtual tubes equals that of the real channel, which implies $\xi_0=1$.
This choice was made in Refs~\onlinecite{minina2014induction} and \onlinecite{minina2015entropic}.
The quantity $\xi$ is a measure of the fraction of the channel cross section effectively 
occupied by polymer \#1, while $\xi_0-\xi$ measures the same for polymer \#2.
Note that the extension length of polymer \#2, $X_2$, is also the distance along the channel 
over which the polymers overlap and, thus, the length of the two virtual tubes. The portion of 
polymer \#1 outside this range is confined solely to the real tube of diameter $D$. Using the 
de~Gennes blob theory (i.e. the extension length of polymer of length $N$ confined to a tube 
of width $D$ is $X\propto ND^{(\nu-1)/\nu}$), it follows that:
\begin{eqnarray}
m_1 D_1^{(\nu-1)/\nu} = N_2 D_2^{(\nu-1)/\nu},
\end{eqnarray}
where $m_1 (\leq N_1)$ is the number of monomers of polymer \#1 that lie within the span of \#2, 
and where the scaling prefactors have canceled. The various quantities are illustrated in 
Fig.~\ref{fig:nestblob}. Using the relation between virtual tube diameters and $D$ introduced 
above, it is easily shown that
\begin{eqnarray}
\xi(m_1/N_2) = \xi_0 \left[1+\left(\frac{m_1}{N_2}\right)^{2\nu/(\nu-1)}\right]^{-1}.
\label{eq:alpha}
\end{eqnarray}

Employing once again the de~Gennes blob theory, the free energy of polymer \#2 is
\begin{eqnarray}
F_2 = N_2D_2^{-1/\nu} = N_2 D^{-1/\nu} (\xi_0-\xi)^{-1/2\nu}.
\label{eq:F2}
\end{eqnarray}
The free energy of polymer \#1 has contributions from the overlapping and
non-overlapping (``overhang'') portions of the chain:
\begin{eqnarray}
F_1 = (N_1-m_1) D^{-1/\nu} + m_1 D^{-1/\nu} \xi^{-1/2\nu}.
\label{eq:F1}
\end{eqnarray}
Note that the scaling prefactors have been omitted from Eqs.~(\ref{eq:F2})
and (\ref{eq:F1}). 

It follows that the free energy in regime~I relative to that of regime~IV is
\begin{eqnarray}
\Delta F =  F_1(m_1) + F_2(m_1) - (N_1+N_2) D^{-1/\nu},
\label{eq:delFtot}
\end{eqnarray}
where the third term is the free energy in regime~IV, where the polymers are
sufficiently separated to not interact. From Eqs.~(\ref{eq:alpha}) -- (\ref{eq:F1}),
it follows that
\begin{eqnarray}
\Delta F(m_1/N_2) = N_2 D^{-1/\nu} f(m_1/N_2;\xi_0),
\label{eq:DelF}
\end{eqnarray}
where
\begin{eqnarray}
f(x;\xi_0) = \xi(x)^{-1/2\nu} x + \left[\xi_0-\xi(x)\right]^{-1/2\nu} - x - 1 ~~~
\label{eq:fxbeta}
\end{eqnarray}
Neglecting fluctuations, the value of $x\equiv m_1/N_2$ is determined by minimization
of $f$. Numerical determination of the minimum yields the linear relation of $x$ and
$\xi$ with respect to $\xi_0$:
\begin{eqnarray}
x \equiv m_1/N_2 & = & 0.79747 + 0.1 \xi_0 \nonumber \\
\xi & = & -0.10903 + 0.53072\xi_0
\label{eq:xalpha}
\end{eqnarray}
in the range of $\xi_0=1-1.4$. For the case of a nested polymer of length $N_2=200$,
this corresponds to values ranging from $m_1=179.51$ and $\xi=0.42348$ for $\xi_0=1$ 
to $m_1=187.51$ and $\xi=0.63576$ for $\xi_0=1.4$. Note that $m_1<N_2$ and 
$\xi > \xi_0-\xi$. This implies that in the overlap regime the nested polymer 
is more longitudinally compressed and occupies a greater fraction of the cross-sectional area than
that of the overlapping portion of the longer polymer.

\section{Nesting free energy function for $\lambda=0$}
\label{app:b}

Figure~\ref{fig:nesting}(a) shows the variation of the free energy with respect to
the extension length difference, $\zeta=X_1-X_2$, for two polymers of similar
contour length and overlapping centers of mass (i.e. $\lambda$=0). In this appendix, we
derive expressions that can be used to account for this behavior. As in Appendix~\ref{app:a},
we follow the approach taken by Minina and Arnold,\cite{minina2014induction,minina2015entropic}
who considered the special case of polymers of equal length. For simplicity, we consider
only the case of $\xi_0=1$.

In Appendix~\ref{app:a}, we considered the case of $\zeta=X_1-X_2>0$. There, the 
conformational free energy can be written as the sum
\begin{eqnarray}
F(m_1) =  F_1(m_1) + F_2(m_1),
\end{eqnarray}
where
\begin{eqnarray}
F_1 = (N_1-m_1) D^{-1/\nu} + m_1 D^{-1/\nu} \xi^{-1/2\nu},
\end{eqnarray}
and
\begin{eqnarray}
F_2 = N_2D_2^{-1/\nu} = N_2 D^{-1/\nu} (1-\xi)^{-1/2\nu}.
\end{eqnarray}
In addition, 
\begin{eqnarray}
\xi(m_1) = \left[1+\left(\frac{m_1}{N_2}\right)^{2\nu/(\nu-1)}\right]^{-1}.
\end{eqnarray}
Note these relations follow from Eqs.~(\ref{eq:F1}), (\ref{eq:F2}) and (\ref{eq:alpha}) for
$\xi_0=1$.  The difference in extension lengths, $\zeta$, is the total span of the portion of 
polymer \#1 located outside the region spanned by \#2. This is given by
\begin{eqnarray}
\zeta(m_1) = (N_1-m_1) D^{(\nu-1)/\nu},
\label{eq:zetaN1p}
\end{eqnarray}
where we omit the prefactor. As both $F$ and $\zeta$ are parameterized
by $m_1$ the relationship between $F$ and $\zeta$ is easily calculated.

Now consider the case where polymer \#1 is nested within polymer \#2, in which case 
$\zeta\equiv X_1-X_2<0$. In addition, $m_1 = N_1$ and $m_2(< N_2)$ are the number of 
monomers of each polymer within the span of \#1. Following the approach described above, 
it is easily shown that
\begin{eqnarray}
\xi(m_2) = \xi_0 \left[ 1 + \left(\frac{N_1}{m_2}\right)^{2\nu/(\nu-1)}\right]^{-1},
\end{eqnarray}
and the contributions to the free energy from both polymers are
\begin{eqnarray}
F_1 = N_1 D^{-1/\nu} \xi^{-1/2\nu}
\end{eqnarray}
and
\begin{eqnarray}
F_2 = (N_2-m_2) D^{-1/\nu} + m_2 D^{-1/\nu}(1-\xi)^{-1/2\nu}.
\end{eqnarray}
Thus, the total free energy,
\begin{eqnarray}
F(m_2) = F_1(m_2) + F_2(m_2),
\end{eqnarray}
is a function of the quantity $m_2$. Since $X_2>X_1$, the extension length 
difference is the {\it negative} of the distance spanned by the portion
of polymer \#2 outside the range spanned by \#1:
\begin{eqnarray}
\zeta(m_2) = -(N_2-m_2) D^{(\nu-1)/\nu}.
\end{eqnarray}
Since both $F$ and $\zeta$ are functions of the variable $m_2$, the relationship
between $F$ and $\zeta$ is once again easily calculated.

Using these relations, which cover both regimes of $\zeta > 0$ and $\zeta < 0$,
we have calculated $F(\zeta)$ for a system with $N_1$=200 and $D=4$ for various
values of $N_2$. The results are shown in Fig.~\ref{fig:nesting}(b).

\section{Additional results illustrating the scaling properties of the free energy}
\label{app:c}

As described in Sec.~\ref{subsec:overlap}, the confined-polymer system is characterized
by four scaling regimes.  In Sec.~IV~D, we argue that the regimes are divided
by boundaries that depend only the values of $N_2/N_1$ and $\lambda/(N_1+N_2)D^{-\beta}$,
where $\beta\approx 0.64$.  This conclusion was made on the basis of the results of 
Fig.~\ref{fig:F.N1=300}. The free energy functions were shown to collapse 
to a universal curve when the axes were scaled such that $F^\prime \equiv F/(N_1+N_2)D^{-\alpha}$ 
is plotted as a function of  $\lambda^\prime\equiv \lambda/(N_1+N_2)D^{-\beta}$, where 
$\alpha=1.82$.  Figure~\ref{fig:F.N1=300}(a) and Fig.~\ref{fig:F.N1=300}(b) show the data 
collapse for fixed $N_1$=300 and fixed $D$=6 for various values of $N_2$. Note that this 
collapse is only expected in regimes~II--IV, and not for regime~I, which accounts for the 
observed deviation from this scaling rule at low $\lambda$. Figure~\ref{fig:F.N1=300}(b)
 and Fig.~\ref{fig:F.N1=300}(d) show the data collapse for fixed $N_1$=300 and fixed $N_2$=200 
for various values of $D$.  As expected, the data collapse in Fig.~\ref{fig:F.N1=300}(d) is 
valid for all regimes, including regime~I, since $N_2$ is fixed.

\begin{figure}[!ht]
\begin{center}
\vspace*{0.2in}
\includegraphics[width=0.80\textwidth]{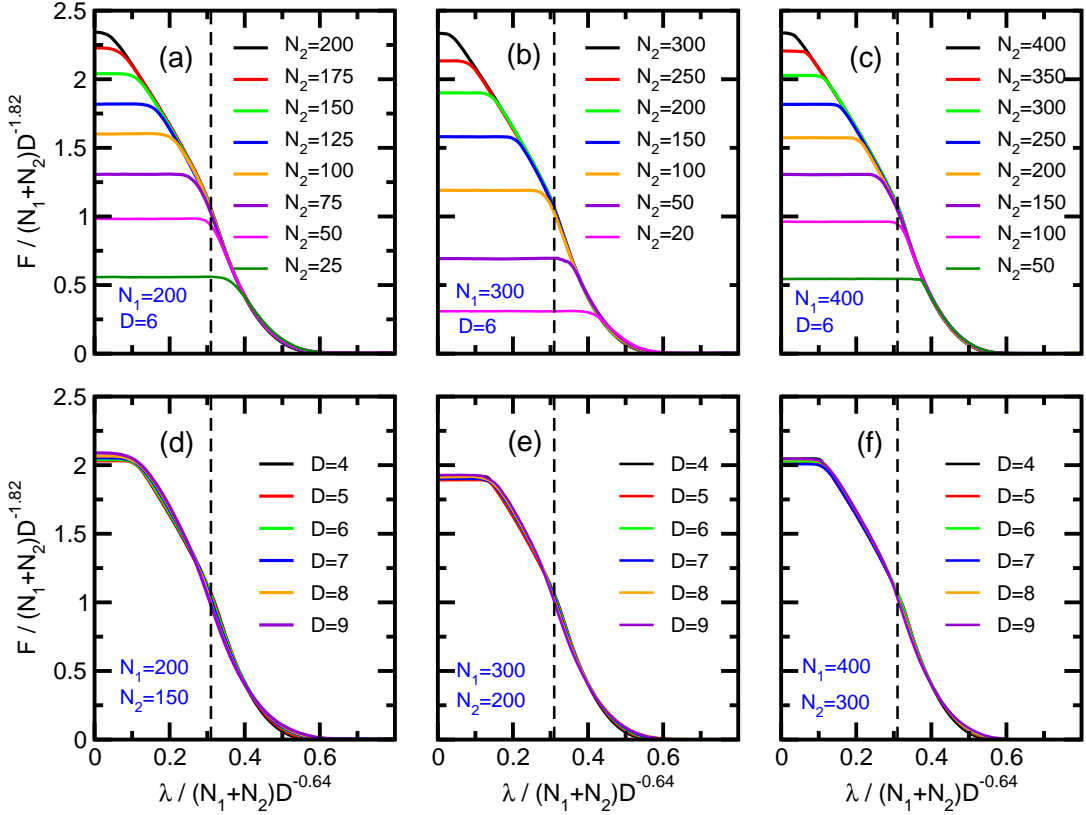}
\end{center}
\caption{Scaled free energy, $F/(N_1+N_2)D^{-1.82}$ versus scaled polymer separation,
$\lambda/(N_1+N_2)D^{-0.64}$. Panels (a) -- (c) each show results for fixed $N_1$
and fixed $D$ for various values of $N_2$. The results are arranged as follows:
(a) $N_1$=200 and $D$=6, (b) $N_1$=300 and $D$=6, and (c) $N_1$=400 and $D$=6.
Panels (d) -- (f) each show results for fixed $N_1$ and $N_2$ for various values of $D$.
The results are arranged as follows: (d) $N_1$=200 and $N_2$=150, (e) $N_1$=300
and $N_2$=200, and (f) $N_1$=400 and $N_2$=300. The vertical dashed line in each
panel marks the approximate boundary between regimes II and III.}
\label{fig:fig11}
\end{figure}

Each of the free energy functions presented in Fig.~\ref{fig:F.N1=300} corresponds to fixed 
$N_1$=300.  The theoretical model predicts that the scaling boundaries in the 
$\lambda^\prime$--$F^\prime$ plane do not depend on the specific value of $N_1$. This is 
equivalent to the prediction that the scaled functions $F^\prime(\lambda^\prime)$ are 
independent of $N_1$. To test this invariance, we have calculated a collection of free energy 
functions for $N_1$=200 and $N_1$=400.  Figure~\ref{fig:fig11} below compares the scaled functions 
calculated for $N_1$=200, 300, and 400.  For each $N_1$, we have varied $N_2$ at fixed $D$ 
(Figs.~\ref{fig:fig11}(a)--(c)), and have also varied $D$ for fixed $N_2$ 
(Figs.~\ref{fig:fig11}(d)--(f)). The expected data collapse in the relevant regimes is 
observed. (Note that this does not include regime~I for Figs.~\ref{fig:fig11}(a)--(c)). 
This result provides further evidence for the veracity of our claim regarding the regime 
boundaries.

\onecolumngrid

\twocolumngrid

\section{Scaling of $F(\lambda)$ in regime~II}
\label{app:d}

In this appendix, we develop a model to help understand the scaling properties
of the free energy function $F(\lambda)$ presented in Sec.~\ref{subsec:scaling}.

Consider the case of two partially overlapping polymers in regime~II. The polymer lengths
are $N_1$ and $N_2\leq N_1$. Generally, the number of monomers of each polymer in the overlap 
region, $m_1$ and $m_2$, need not be equal. However, we note from Fig.~\ref{fig:regime_I_scale}(a)
that these are indeed very close to equal in this regime. Consequently, we define $m\equiv m_1=m_2$.
In addition, we choose the diameters of the virtual tubes to be $D_1=D_2=D/\sqrt{2}$. This corresponds
to choosing $\xi = \frac{1}{2}$ and $\xi_0=1$ in Appendix~\ref{app:a}. The choice $\xi=\frac{1}{2}$
corresponds to the observation that $m_1=m_2$.  Choosing $\xi_0$=1 corresponds to not considering 
lateral interpenetration of the chains. We make this simplification since choosing instead $\xi_0>1$ 
changes only the scaling prefactors and not the functional form of the scaling nor the values of
the exponents. 

The free energy of a partially overlapping chain is given by
\begin{eqnarray}
F_1 & = & c_0(N_1-m) D^{-1/\nu} + c_0(N_2-m)D^{-1/\nu} \nonumber \\
    &   & + 2 c_0 m (D/\sqrt{2})^{-1/\nu},
\end{eqnarray}
where $c_0$ is a scaling constant.  The first and second terms account for the parts of each 
polymer outside the overlap region. The third term is the contribution of the overlapping
portion of the two polymers. The free energy of two polymers that are far apart (i.e. in regime~IV)
is given by
\begin{eqnarray}
F_2 = c_0N_1 D^{-1/\nu} + c_0N_2D^{-1/\nu}.
\end{eqnarray}
The free energy in regime~II relative to that of regime~IV is thus
\begin{eqnarray}
F = F_1-F_2 = c_0^\prime m D^{-1/\nu},
\end{eqnarray}
where $c_0^\prime \equiv (2^{1+1/2\nu}-2)c_0$. It follows that
\begin{eqnarray}
\frac{F}{(N_1+N_2)D^{-1/\nu}} = c_0^\prime\left(\frac{m}{N_1+N_2}\right)
\label{eq:FN1N2D}
\end{eqnarray}

To determine the relationship between the free energy and $\lambda$, we need to find the relation
between $m$ and $\lambda$. For convenience, choose $z=0$ to be the center of the overlap
region, and define $z_1(<0)$ to be the center of mass of polymer \#1 and $z_2(>0)$ to be
the center of mass of polymer \#2. It follows that
\begin{eqnarray*}
z_1 = \left(\frac{N_1-m}{N_1}\right) \left(-{\textstyle\frac{1}{2}}L_{\rm ov} 
-{\textstyle\frac{1}{2}}L_1\right),
\end{eqnarray*}
where $L_1$ is the extension length of polymer \#1 outside of the overlap region. 
Note that $L_1 =c(N_1-m)D^{1-1/\nu}$, where $c$ is a constant, and also that
$L_{\rm ov}=c^{\prime} m D^{1-1/\nu}$, where $c^{\prime}\equiv 2^{-1/2+1/2\nu}c$.
It can then be shown that
\begin{eqnarray}
-2z_1/D^{1-1/\nu} & = & (c^{\prime\prime}-c) m + cN_1 - c^{\prime\prime} m^2/N_1
\label{eq:z1}
\end{eqnarray}
where $c^{\prime\prime}\equiv c^\prime - c$. Likewise, it can be shown that $z_2$ is given by
\begin{eqnarray}
2z_2/D^{1-1/\nu} & = & (c^{\prime\prime}-c) m + cN_2 - c^{\prime\prime} m^2/N_2
\label{eq:z2}
\end{eqnarray}
From Eqs.~(\ref{eq:z1}) and (\ref{eq:z2}) it follows that $\lambda\equiv z_2-z_1$ satisfies:
\begin{eqnarray}
\frac{\lambda}{(N_1+N_2)D^{1-1/\nu}} & = & \Delta c\left(\frac{m}{N_1+N_2}\right) + {\textstyle\frac{1}{2}}c \nonumber \\
                                     && - {\textstyle\frac{1}{2}} c^{\prime\prime} \left(\frac{m}{N_1+N_2}\right)^2
                                             \frac{(N_1+N_2)^2}{N_1N_2}, \nonumber \\
&&
\label{eq:lamscale}
\end{eqnarray}
where $\Delta c \equiv c^{\prime\prime}-c$.

\begin{figure}[!ht]
\begin{center}
%\vspace*{0.2in}
%\includegraphics[width=0.4\textwidth]{figures/lambda.eps}
\includegraphics[width=0.4\textwidth]{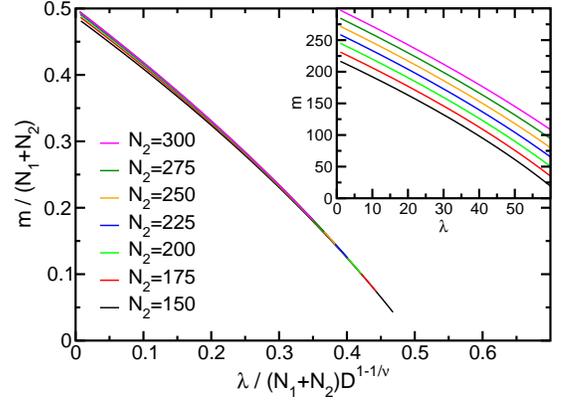}
\end{center}
\caption{Variation of $m/(N_1+N_2)$  with $\lambda/(N_1+N_2)D^{1-1/\nu}$ obtained by solving 
Eq.~(\ref{eq:lamscale}) for the case of $N_1$=300 and $D$=6. The inset shows the unscaled data.  }
\label{fig:lambda}
\end{figure}

In the special case where $N_1=N_2\equiv N$, the factor $(N_1+N_2)^2/N_2N_2$ reduces to a constant,
and the quadratic equation for the variable $m/(N_1+N_2)\propto m/N$ can be solved to yield a
relation of the form $m/N = u(\lambda/ND^{1-1/\nu})$, as noted in Ref.~\onlinecite{polson2018segregation}.
Substitution into Eq.~(\ref{eq:FN1N2D}) then yields a scaling of $F(\lambda)$ consistent in form
with that seen for the data presented in Sec.~\ref{subsec:scaling}; that is,
$F(\lambda;N,D) = ND^{-1/\nu} f(\lambda/ND^{1-1/\nu})$, where $f(x) = u(x)$ in regime~II. However, 
in the general case that $N_1\neq N_2$, the resulting expression for $m/(N_1+N_2)$ has a residual 
dependence on $N_1$ and $N_2$ beyond that implicit in the dependence on scaled variable 
$\lambda/(N_1+N_2)D^{1-1/\nu}$. Consequently, the theoretical model does not
exactly yield the scaling described by Eq.~(\ref{eq:Fscale}). In practice, however, the quantitative
effect of the factor $(N_1+N_2)^2/N_2N_2$ is minimal. Figure~\ref{fig:lambda} shows results for
$m/(N_1+N_2)$ vs $\lambda/(N_1+N_2)D^{1-1/\nu}$ for $N_1$=300 and $D$=6. The curves for various 
$N_2$ all come close to collapsing to a universal curve. The greatest deviation occurs for low
$\lambda$ for cases where the system has transitioned into regime~I, where these results are not
relevant. Similar results were obtained for other values of $N_1$. We conclude that the theoretical
model produces results that come very close to yielding the predicted scaling of Eq.~(\ref{eq:Fscale})
in regime~II.

% \bibliography{paper}
% \bibliographystyle{apsrev4-1}


\end{document}